\documentclass{article}
\usepackage{leap96}
\newcommand{\be}{\begin{equation}}
\newcommand{\ee}{\end{equation}}
\title{\Large Hybrid mesons: old prejudices and new spectroscopy}
\author{$\rm Yu.S.Kalashnikova^a$}
\date{${}^a$Institute of Theoretical and Experimental Physics,
117259, Moscow, Russia}
\begin{document}
\twocolumn[
\maketitle
\begin{abstract}
\hspace*{1em} The models for hybrid mesons are discussed, in which the gluonic 
excitations manifest themselves as the vibrations of the quark-
antiquark QCD string. The predictions for the spectra, decays and
mixing with hadronic channels are presented.
\end{abstract}
]

{\parindent=0cm \bf 1. INTRODUCTION}\\

The problem of existence or nonexistence of hadrons with gluonic
content attracts a lot of attention both from theoretical and 
experimental point of view. Rather general arguments from QCD tell 
that the theory, even quenched in quarks, should possess nontrivial
spectrum, while the present state--of--art does not allow to derive it 
from first principles. A quite respectable way to deal with such
situation is to try to rewrite the generating functional of QCD in
nonperturbative region in terms of constituent degrees of freedom. To
formulate a string theory is, apparently, a suitable way to do it.

As the QCD is not a string theory, a question arises: is it possible to
introduce the notion of constituent glue in this effective string theory.
To answer this question it is instructive to look for experience from
more simple theories, namely the two--dimensional QCD in the large
$N_c$ limit. The ${\rm QCD}_2$ with fundamental fermions was analysed long
ago [1], and it was found that the spectrum of the model is exhausted
by $q \bar q$ meson states joined by the string. These states lie on the
Regge trajectory\\

{\parindent=0cm
$
M_n^2 = n g^2 N_c\hspace*{5cm}(1)
$\\

with the density of states rising linearly with the mass, 
$dn/dM\sim M$.
The simple spectrum (1) is completely due to the fact that there is no
dynamical gluons in the ${\rm QCD}_2$, and the Coulomb force is confining
in two dimensions.

}

In contrast to t'Hooft model, the ${\rm QCD}_2$ with fermions which belong
to the adjoint representation of the colour group is less trivial even
in the large $N_c$ limit. It happens because the presence of adjoint
matter allows for more complicated string configurations. It was shown
by explicit calculations [2] that the number of Regge trajectories is
infinite, and the density of states grows exponentially with the mass.
Moreover, the very notion of trajectory remains reasonable only for the
lowest states; with the increase of the mass more and more states from
different trajectories enter the game, the trajectories begin to overlap
and the spectrum becomes rather stochastic.

It is the same behaviour as one expects in the ${\rm QCD}_4$, with dynamical
gluons playing the role of adjoint matter. The effective string model 
should be arranged to allow the gluons to populate the string and to be
responsible for complicated string configurations.\\

{\parindent=0cm \bf 2. SPECTRUM ESTIMATIONS}\\

The area law asymptotics for the Wilson loop conveniently provides us
with the action of the string, and naive expectations from the string
spectrum read\\
\smallskip

{\parindent=0cm
$
M^2 = 2 \pi \sigma (L + 2\nu)\hspace*{4.2cm}(2)
$\\
\smallskip

for the string Regge trajectory, were $L$ is the orbital momentum, $\nu$
is the vibrational quantum number, and $\sigma$ is the string tension. 
Placing quarks at the ends of the string we have new degrees of freedom, so
that the Lagrangian of the system becomes\\
\smallskip

$
L = - m_1\sqrt{\dot{x}_1^2} - m_2\sqrt{\dot{x}_2^2} + L_{\rm string}\;.
\hspace*{2cm}(3)
$\\

}

The system (3) was extensively studied in [3] in the so-called straight-line
approximation which corresponds to the special kind of the string motion 
with frozen vibrations ($\nu = 0$). In the limit of large quark masses the
Lagrangian (3) gives rise to the effective quasipotential Hamiltonian
for the $q \bar q$ system, \\

{\parindent=0cm
$
H=\sqrt{p^2+m^2_1}+\sqrt{p^2+m^2_2}+\sigma r,\hspace*{2cm}(4)
$\\

and for high excitations the Regge trajectory
takes the form\\

}

\smallskip

{\parindent=0cm
$
M^2 = 2 \pi \sigma (l + 2n_r) + {\rm quark\;mass\;corrections}.\;(5)$\\
\smallskip

}

It is natural to identify the $q \bar q$ system connected by straight - line
string with conventional $q \bar q$ meson, while the string vibrations are
responsible for the gluonic excitations of the QCD string, forming hybrids,
with Regge trajectory given by\\
\smallskip

{\parindent=0cm
$
M^2 = 2 \pi \sigma (L + 2n_r + 2\nu).\hspace*{3.1cm}(6)
$\\

}

The lower part of the spectrum (6) is distorted by quark mass corrections, and
the actual pattern of distortion depends on the model adopted (see e.g the
results from the flux tube model [4] and the constituent string model [5]).
It is clear that for hybrids with light quarks there is only one dimensional
parameter, $\sqrt\sigma$, and, in contrast to nonrelativistic systems, 
there is no gap between radial/orbital/vibrational excitations in the 
$q \bar q$ system. Quark mass corrections, short range effects and mixing
make the trajectories given by oversimplified estimation (5) overlap, 
and
the spectrum is expected to be rather stochastic (like in the 
${\rm QCD}_2$ with
adjoint fermions?).\\

{\parindent=0cm \bf 3. HYBRIDS DECAYS}\\

The existing models [4,5] agree that the lowest $q \bar q g$ hybrids
have the masses around 1.7 -- 1.9 Gev, ie in the mass region populated
by the radial and orbital $q \bar q$ mesons. There is also the agreement 
concerning quantum numbers of the P--odd part of hybrid spectrum:\\
\smallskip

{\parindent=0cm
$
J^{PC} = 0^{-+},\; 1^{-+},\; 2^{-+},\; 1^{--}.
$\\
\smallskip

}

It appears that just the same quantum numbers cause a lot of interest
in "hybrid hunting".

Indeed, it was established in a lot of ways [6--8] that the famous selection 
rule exists for hybrid decay: the ground state hybrid does not decay into
two ground state mesons. More accurate statement is that the decay
matrix element vanishes for two--body final states with identical space
wave functions. It means that the main decay modes of a hybrid are the 
modes involving S--wave ground state meson +  P--wave meson. There is also 
less famous and more model dependent selection rule [7,8]: not only the total
angular momentum, but also the total spin of constituents is conserved in
the decay process. Application of the selection rules to the ground state
hybrid decays leads to the following list of favoured decay modes:\\

{\parindent=0cm
Isovectors:\\
\smallskip

$1^{--}\to  \pi a_1,\; \rho \varepsilon$\\
\smallskip

$0^{-+}\to  \pi\; \varepsilon,\; \pi f_0$\\
\smallskip

$1^{-+}\to  \pi b_1,\; \pi f_1$\\
\smallskip

$2^{-+}\to  \pi f_2$\\

Isoscalars:\\
\smallskip

$1^{--}\to  \omega \varepsilon,\; \omega f_0$\\
\smallskip

$0^{-+}\to  \eta \varepsilon,\; \pi a_0$\\
\smallskip

$1^{-+}\to  \pi a_1$\\
\smallskip

$2^{-+}\to  \pi a_2,\; \eta f_2$\\

}

An additional advantage exists which can help to tell hybrids from 
radially excited quarkonia: because of the node in the wave function of
radial $q \bar q$ excitation the decay of $nS$ $q \bar q$ into S--wave
ground state meson + P--wave meson is suppressed. Recently the
comprehensive analysis of higher quarkonia decays in the ${}^3P_0$ model
was performed [9], and it was stressed once more that the studies of the 
decays into both S + S and S + P mesons can provide useful information on
quarkonium or hybrid assignements. Indeed, the mass estimations are model
dependent, and are to be supplied by other considerations, like "extra"
states, exotic quantum numbers and so on. The decay signatures play an
important role, and the existing hybrid candidates were isolated basing
on the analysis of the decay modes mainly. 
 
It is well known that something is wrong with $\rho(1450)$ as the $2S$
radial $q \bar q$: not only the four--pion mode is dominant, but also the
$e^+e^- \to  \pi^+ \pi^- \pi^+ \pi^-$ crossection is several times larger 
than $e^+e^-\to   \pi^+ \pi^- \pi^0 \pi^0$ around this resonance [10]. In 
terms of two - body intermediate states it means that $\pi a_1$ is the
main decay channel, and $\pi h_1$ is remarkably weak, in accordance with 
expectations for hybrid (spin conservation selection rule). On the other 
hand, the electromagnetic coupling of a hybrid is much smaller than of
$2S\; q \bar q$, which suggests the mixing scheme [11] to describe
$\rho(1450)$.

The $1^-\;(0^{-+})$ pion with the mass 1.8 GeV was seen in [12], and 
rediscovered 
by VES [13]. It decays into $\pi f_0$ and $\pi K \bar K$, and has a moderate
width of about 200 MeV, in contrast to expectations for $3S\; q \bar q$ pion.
As the $3S$ state should be placed somewhre around, careful analysis of
other modes, especially $\rho \omega$ [9] is needed to confirm the hybrid
component.

The $1^{-+}$ channel was always thought of as a hybrid one, because such
quantum numbers are unaccessible in the $q \bar q$ sector. The resonant
activity in $\pi f_1$ $1^{-+}$ wave [14] is rather promising; it should be
noted, however, that there is another mode, $\pi b_1$, with the ratio of
branchings $\pi b_1 / \pi f_1 \sim 4$ estimated for a hybrid [8]. The 
observation of $\pi b_1$ signal would provide a strong support in favour
of hybrid interpretation. It worth mentioning that BNL has reported
the resonances in this wave with much lower masses [15]; if confirmed,
this observation might change the adopted hybrid picture in a very
funny way.

The tensor channel $2^{-+}$ is interesting {\it per se}. The $\pi_2(1670)$ 
is a
respectable member of $q \bar q$ ${}^1D_2$ nonet, with strong coupling
to $\pi f_2$ and $\rho \pi$. There are also evidences for other
tensors in the same mass region, with the same decay pattern [16,17].
The absence of isoscalar $2^{-+}$ states was always intriguing, and, at last,
$\eta_2$ states begin to appear: the $\eta_2(1875)$ from Crystall Ball [18]
and $\eta_2(1645)$ and $\eta_2(1875)$ from Crystall Barrel [19]. As the
$q \bar q$ in this wave is orbitally excited with no suppression of
S--wave + P--wave mesons, the large isovector $\pi f_2$ or isoscalar
$\pi a_2$ modes do not provide decisive arguments for hybrid assignment,
and further studies are needed to identify tensor hybrids. Actually, this
channel is a perfect playing ground to study the $q \bar q$--$q \bar q g$
mixing.\\

{\parindent=0cm \bf 4. HADRONIC SHIFTS}\\

The first results of hybrid searches look rather promising, but the life
is not as simple as it might seem. The mechanism exists which might change
the simple constituent picture drastically, and this mechanism is mixing
with hadronic channels and unitarity effects.

The necessity to perform the coupled channel analysis in the formalism 
which respects unitarity was advocated for long by N.Tornqvist. The
unitarity effects can be included via dispersion relations approach, and
the scalar nonet was recently analysed in such way [20]. The physical mass
of the resonance appears to be considerably shifted with respect to the
bare constituent model mass due to the coupling to hadronic channels.
Correspondingly, the wave function of the physical state contains
considerable contribution of hadronic molecule component. The magnitude
of effect depends not only on the strength of coupling, but also on
how close to the threshold the resomance is, and on the partial wave
of hadronic final state. There even might be no one--to--one
correspondence between bare state and resonance; the same bare state
can manifest itself as relatively narrow structure and profound
threshold effect at the same time. Such situation usually occurs if
the final hadrons are in relative S--wave, as it takes place for
dominant hybrid decays into S--wave + P--wave mesons.

Indeed, the physical mass $m$ of the resonance is defined as the pole
of the hadronic $S$--matrix\\

{\parindent=0cm
$
m^2 = m_0^2 + \Delta,\;\;\;\;\Delta =\sum_i\Delta (i),\hspace*{2.15cm}(7)
$\\

where $m_0$ is the bare constituent mass, and the hadronic shift $\Delta$
is the sum over contributions from all hadronic channels $i$. The 
imaginary part of $\Delta(i)$ is given by decay amplitude of the bare
state into channel $i$, and the real part is calculated via dispersion
relation:\\

$
{\rm Re}\Delta(i) =\displaystyle\int_{\displaystyle s_i}^{\displaystyle
\infty}ds' 
\frac{\displaystyle{\rm Im}\Delta(s')}{\displaystyle{s - s'}}.
\hspace*{2.8cm}(8)
$\\

If the resonance is far from the threshold, one may neglect the dependence
on $s$ in the denominator of (8). The approximate sum rule [21] was
established for the hadronic shift far from the threshold, which actually
means that effect of hadronic loops is simply to renormalize the
bare mass.

}

This sum rule is obviously strongly violated by nearby thresholds, and 
S--wave thresholds are the strongest ones: the imaginary part ${\rm Im}\Delta$
behaves as $k_i^{2l+1}$, and for $l = 0$ one has\\

{\parindent=0cm
$
\frac{\displaystyle\partial}{\displaystyle\partial s}
{\rm Re}\Delta(i)=-\displaystyle\int_{\displaystyle s_i}^{\displaystyle
\infty}ds'
\frac{\displaystyle{\rm Im}\Delta(s')}{\displaystyle(s-s')^2}
\to\infty\hspace*{1.1cm}(9)
$\\
at the threshold.
}

Such cusp-like behaviour of the hadronic $S$-matrix might have a drastic
effect on observables, obscuring the interpretation of data in terms
of bare states. Moreover,the strength of the cusp might be smoothened
over some mass interval in the case of hybrid decay into S--wave +
P--wave mesons, because the P--wave mesons are usually broad, and it
will confuse the picture even more.\\
      
{\parindent=0cm \bf 5. CONCLUSIONS}\\

To summarise, a lot of evidence already exists that the mesons in
the mass region 1.5 --2.0 GeV contain admixture of constituent glue.
Still the final conclusions should be done only after performing the
coupled channel analysis of existing data.\\     

\parindent=0cm {\bf REFERENCES}\\

1. \hspace*{0.2cm}G.t'Hooft, Nucl.Phys., {\bf B75}, (1974) 461\\
2. \hspace*{0.2cm}S.Dalley,~I.R.Klebanov,~Phys.Rev.,~{\bf D47}\\ 
\hspace*{0.5cm} (1993) 2517\\
3. \hspace*{0.2cm}A.Yu.Dubin, A.B.Kaidalov, Yu.A.Simonov, \\
\hspace*{0.5cm} Phys.Lett., {\bf B323} (1994) 41\\
4. \hspace*{0.2cm}T.Barnes, F.E.Close, E.S.Swanson, Phys.\\
\hspace*{0.5cm} Rev., {\bf D52} (1995) 5242\\
5. \hspace{0.2cm}Yu.S.Kalashnikova,~Yu.B.Yufryakov,~Phys.\\ 
\hspace*{0.5cm} Lett., {\bf B359} (1995) 175\\
6. \hspace*{0.2cm}N.Isgur, R.Kokoski, J.Paton, Phys.Rev.Lett.,\\ 
\hspace*{0.5cm} {\bf 54} (1985) 869\\
\hspace*{0.5cm} F.E.Close,~P.R.Page,~Nucl.Phys.~{\bf B443}~(1995)\\
\hspace*{0.5cm} 233\\
7. \hspace*{0.2cm}A.Le~Yaonanc,~L.Oliver,~O.Pene,~J.-C.Ray- \\
\hspace*{0.5cm} nal, Z.Phys. {\bf C28} (1985) 309\\
8. \hspace{0.2cm}Yu.S.Kalashnikova, Z.Phys., {\bf C62} (1994) 323\\
9. \hspace{0.2cm}T.Barnes, in these Proceedings\\
\hspace*{0.5cm} T.Barnes, F.E.Close, P.R.Page,E.S.Swanson,\\ 
\hspace*{0.5cm} MC--TH-96/21, 
ORNL--CTP--96--09, RAL--\\
\hspace*{0.5cm} 96--039, hep--ph/9609339\\  
10. A.B.Clegg, A.Donnachie, Z.Phys., {\bf C62}\\
\hspace*{0.5cm} (1994) 455\\
11. A.Donnachie, Yu.S.Kalashnikova, Z.Phys.,\\
\hspace*{0.5cm} {\bf C59} (1993) 621\\
12. G.Bellini {\it et.al.}, Phys.Rev.Lett., {\bf 48} (1982) \\
\hspace*{0.5cm} 1697\\
13. D.Ryabchikov, in Proceedings of Hadron'95,\\ 
\hspace*{0.5cm} Manchester, UK, 1994\\
14. J.H.Lee {\it et.al.}, Phys.Lett. {\bf B323} (1994) 227\\
\hspace*{0.5cm} S.U.Chung, in these Proceedings\\
15. S.U.Chung, in these Proceedings\\
16. G.T.Condo {\it et.al.}, Phys.Rev., {\bf D43} (1991)\\ \hspace*{0.5cm} 
2787\\
17. C.Daum {\it et.al}, Nucl.Phys., {\bf B182} (1981) 269\\
18. K.Karch {\it et.al.}, Z.Phys., {\bf C54} (1992) 331\\
19. D.V.Bugg, in these Proceedings\\
\hspace*{0.5cm} D.V.Bugg {\it et.al.}, Z.Phys.C, in press \\
20. N.A.Tornqvist, Z.Phys., {\bf C68} (1995) 647\\
21. P.Geiger, N.Isgur, Phys.Rev., {\bf D44} (1991)\\ 
\hspace*{0.6cm}799
\end{document}